\documentclass[aps,prl,twocolumn,amsmath,amssymb,groupedaddress,longbibliography]{revtex4-1}
\usepackage[english]{babel}

\usepackage{amssymb}
\usepackage{dcolumn}
\usepackage{bm}
\usepackage{graphicx}
\usepackage{amsmath}

\usepackage{graphicx}
\graphicspath{{pict/}{}}

\usepackage{dcolumn}
\usepackage{bm}
\usepackage[pdfstartview=FitH, CJKbookmarks=true, bookmarksnumbered=true, bookmarksopen=true, colorlinks=true, pdfborder=001, citecolor=blue, linkcolor=blue, linktocpage=true] {hyperref}

\begin{document}

\preprint{APS/123-QED}
\title{Shortcuts to adiabatic Thouless pumping}

\author{Wenjie Liu $^{1,2,3}$}

\author{Yongguan Ke $^{2,3,1,4}$}
\altaffiliation{Email: keyg@mail2.sysu.edu.cn}

\author{Chaohong Lee $^{2,3,1}$}
\altaffiliation{Email: chleecn@szu.edu.cn}

\affiliation{$^{1}$Quantum Science Center of Guangdong-Hongkong-Macao Greater Bay Area (Guangdong), Shenzhen 518045, China}

\affiliation{$^{2}$Institute of Quantum Precision Measurement, State Key Laboratory of Radio Frequency Heterogeneous Integration, Shenzhen University, Shenzhen 518060, China}

\affiliation{$^{3}$College of Physics and Optoelectronic Engineering, Shenzhen University, Shenzhen 518060, China}

\affiliation{$^{4}$Laboratory of Quantum Engineering and Quantum Metrology, School of Physics and Astronomy, Sun Yat-Sen University (Zhuhai Campus), Zhuhai 519082, China}


\date{\today}

\begin{abstract}

Thouless pumping, the quantized transport of particles in a cyclic adiabatic evolution, faces a challenge: slow driving may exceed the coherent time, while fast driving may break quantization.
To address this dilemma, we propose to speed up Thouless pumping using shortcuts to adiabaticity.
By using counterdiabatic theory, we analytically derive the controlled Hamiltonian for implementing dispersion-suppressed Thouless pumping beyond the adiabatic regime.
Compared to traditional Thouless pumping methods, our fast topological pumping approach offers remarkable advantages.
Firstly, it enables a substantial reduction of pumping time up to $11$ orders of magnitude faster than the traditional approach.
Secondly, our method effectively suppresses wavepacket diffusion, further enhancing its efficiency.
Furthermore, we demonstrate the resilience of our protocol against moderate noise levels.
Our study offers a practical and efficient method for achieving fast topological pumping beyond the adiabatic regime.

\end{abstract}

\maketitle

{\em Introduction}. Thouless pumping, a phenomenon of quantized particle transport within periodically modulated potentials~\cite{1983PhysRevB276083,1984JPAMG,RevModPhys821959,2023TPandTopo}, requires uniform occupation of an isolated band and adiabatic evolution.
It has been successfully demonstrated in experiments involving ultracold atoms~\cite{2016NP12296,2016NP12350,2016PhysRevLett117170405,2018N55355}, photons~\cite{2012PhysRevLett109106402,2018N55359,2020LSA9178} and spins~\cite{2018PhysRevLett120120501}.
%
%
To overcome the need for uniform band occupation, an alternative approach is to introduce a tilted field that allows efficient sampling of the entire Bloch band through Bloch oscillations~\cite{2020PhysRevResearch2033143,2021PhysRevB104104314,2023PhysRevResearch5013020}.
However, the challenge lies in taking extended duration to achieve adiabatic evolution~\cite{2018PhysRevLett120106601}, which poses difficulties in maintaining quantum coherence during realistic experiments and causes significant diffusion of the wavepacket~\cite{ke2016topological,PhysRevB.100.064302}.
These limitations hamper applications of Thouless pumping in quantum information processes, such as quantum state transfer.
Several methods have been explored to speed up Thouless pumping, including the introduction of dissipation~\cite{2020JStatMech,2020NatC113758} and correlations~\cite{2021PhysRevA104063315,2022SciPostPhys126203}.
However, these approaches only provide marginal increases in the driving frequency, as they are limited by the energy gap.

Shortcuts to adiabaticity (STA)~\cite{UNANYAN199748,PhysRevLett104063002,2010PhysRevLett105123003,AAMOP621172013,RevModPhys91045001} provide a flexible framework for accelerating quantum dynamics by manipulating time-dependent Hamiltonians.
These techniques aim to achieve the desired evolution in a much shorter time.
The counterdiabatic approach~\cite{JPhysChemB2005,Berry2009} specifically allows for a swift and seamless steering of a quantum system along a predetermined adiabatic trajectory, eliminating transitions.
STA protocols have been successful in accelerating quantum control~\cite{PhysRevLett104063002,2010PhysRevLett105123003,NP20118147,PhysRevLett2013110240501,NC201671,NP201713330} and quantum computing~\cite{ConorMcarXiv2023}, accelerating particle transport in phase space~\cite{NC2016712999}, and generating entangled states for high-precision measurements~\cite{2012PhysRevA86063623,2012PhysRevA86023615,2018NJP20015010,2018NJP20055009}.
When combined with topological physics, STA becomes a powerful tool for efficiently measuring the Berry phase~\cite{PhysRevA95042345}, simulating topological phase transitions~\cite{SciChinaPhysMechAstron2018}, and understanding non-quantized geometric pumping~\cite{2020PhysRevLett124150603,2020PhysRevLett124150602}.
Although STA has been widely explored in few-level systems,  it is still unclear how to extend to multi-level topological systems.
A significant and captivating challenge remains in accelerating quantized Thouless pumping using STA.

In this Letter, we analytically derive the time-dependent controlled Hamiltonian for achieving Thouless pumping with STA techniques by considering the prototypical Rice-Mele model in momentum space~\cite{1982PhysRevLett491455}.
This approach aims to achieve rapid Thouless pumping beyond the limitations of the adiabatic regime.
Using STA, we can enhance pumping efficiency while maintaining quantization in the accelerated Thouless pumping process.
Surprisingly, the wavepacket width can be suppressed due to a shorter evolution time.
Our STA Thouless pumping method demonstrates superior efficiency and exhibits resilience against Hamiltonian noises.
To provide a comprehensive comparison, we analyze the dynamics of wavepackets in real space for both traditional Thouless pumping and STA Thouless pumping.
When the adiabatic condition is satisfied for both traditional Thouless pumping and STA Thouless pumping, the diffusion of wavepacket can be prominently suppressed in STA Thouless pumping, compared to traditional Thouless pumping.
Additionally, we highlight the feasibility of implementing our STA topological pumping in accessible ultracold atomic systems.

{\em Model and formalism}. We consider the Rice-Mele model in momentum space, which is described by the Hamiltonian matrix
\begin{equation}\label{ChoosenModel}
\hat{H}_{0}(k, t)=\left(\begin{array}{cc}
h^z_0 & h^x_0-i h^y_0 \\
h^x_0+i h^y_0 & -h^z_0
\end{array}\right)
\end{equation}
with $\left( h^x_0, h^y_0, h^z_0 \right)$ = $\left( 2 J \cos (k), 2 \delta_0 \sin (k)\sin (\phi), \Delta_0 \cos (\phi) \right)$ and $\phi=\phi_0+\omega t$.
Its corresponding position-space form belongs to a sublattice structure where the length of unit cell is $d=2a$~\cite{SM}.
For convenience, we set $a=1$ and $\hbar=1$ as dimensionless quantities.
$J$, $\delta_0$, and $\Delta_0$ are the hopping strength, the modulation strength of hopping, and the modulation strength of onsite-energy, respectively.
$k$ is a quasimomentum in the first Brillouin zone $[0,2 \pi / d]$ and $\omega$ is the modulation frequency corresponding to the modulation period $T_m=2\pi/\omega$.
There exists a translational symmetry in the time domain with $\hat{H}_{0}(k, t)=\hat{H}_{0}(k, t+T_m)$.
All eigenstates of Hamiltonian~\eqref{ChoosenModel} return to themselves over a period $T_m$.
Through solving the eigenequation, the eigenstates can be exactly given as
\begin{equation}\label{EigenStates}
\left|\lambda_{+}\right\rangle=\left(\begin{array}{c}
\cos \frac{\theta}{2} \\
\sin \frac{\theta}{2} e^{i \varphi}
\end{array}\right), \left|\lambda_{-}\right\rangle=\left(\begin{array}{l}
-\sin \frac{\theta}{2} e^{-i \varphi} \\
\cos \frac{\theta}{2}
\end{array}\right)
\end{equation}
with eigenvalues being $\varepsilon_{\pm}=\pm \sqrt{(h^x_0)^2+(h^y_0)^2+(h^z_0)^2}$.
The spherical angles $\theta \in[0, \pi)$ and $\varphi \in[0,2 \pi)$ are defined as
\begin{equation}
\cos \theta=\frac{h^z_0}{\varepsilon_{+}}, \quad e^{i \varphi}=\frac{h^x_0+i h^y_0}{\sqrt{(h^x_0)^2+(h^y_0)^2}}.
\end{equation}

According to counterdiabatic theory~\cite{Berry2009}, we obtain the controlled Hamiltonian
\begin{equation}\label{ControlH}
\hat{H}_C(k,t)=\left(\begin{array}{cc}
\frac{1-\cos ^2 \theta}{2} \dot{\varphi} & \left(-i \frac{\dot{\theta}}{2}-\frac{\sin 2 \theta \dot{\varphi}}{4} \right) e^{-i \varphi} \\
\left(i \frac{\dot{\theta}}{2}-\frac{ \sin 2 \theta \dot{\varphi}}{4}\right) e^{i \varphi} & -\frac{1-\cos ^2 \theta}{2} \dot{\varphi}
\end{array}\right),
\end{equation}
see~\cite{SM} for details of  derivation and gauge invariance.
Subsequently, the total Hamiltonian becomes $\hat{H}(k, t)=\hat{H}_0(k, t)+\hat{H}_C(k, t)$, which can be effectively simulated via a three-level ultracold system~\cite{NC201671}.
For a gapped Hamiltonian $\hat{H}_0(k, t)$, its controlled Hamiltonian $\hat{H}_C(k, t)$ guarantees that if the initial state is prepared as one of eigenstates of Hamiltonian $\hat{H}_0(k, t)$, then it will follow the corresponding instantaneous eigenstate even for a much higher driven frequency.
For a two-band Rice-Mele model $\hat{H}_0(k, t)=\vec{h}_0 \cdot \vec{\sigma}$,
the Chern number is defined as a two-dimensional closed-surface integral
\begin{equation}
C=\frac{1}{2 \pi} \int_0^{2 \pi / d} \int_0^{T_m} \mathcal{F}_0 d k d t,
\end{equation}
where the Berry curvature is $\mathcal{F}_0=\left(\partial_k \tilde{h}_0 \times \partial_\tau \tilde{h}_0\right) \cdot \tilde{h}_0/2$ and the unit vector of the effective magnetic field is $\tilde{h}_0=\vec{h}_0 / |\vec{h}_0|=\vec{h}_0 / \varepsilon_{+}$~\cite{RevModPhys821959}.

The evolved state $|\psi(k,t)\rangle$ obeys the Schr\"{o}dinger equation $i \hbar \frac{\partial}{\partial t}|\psi(k,t)\rangle=\hat{H}(k, t)|\psi(k,t)\rangle$.
Once the controlled Hamiltonian $\hat{H}_C(k, t)$ guarantees the state $|\psi(k,t)\rangle$ evolves along the instantaneous eigenstate of $\hat{H}_0(k, t)$, we have
$\tilde{h}_0=\tilde{\sigma}=-\left(\left\langle\hat{\sigma}_x\right\rangle,\left\langle\hat{\sigma}_y\right\rangle,\left\langle\hat{\sigma}_z\right\rangle\right)$ with $\langle\hat{\sigma}_{x,y,z}\rangle=\langle\psi(k, t)|\hat{\sigma}_{x,y,z}| \psi(k, t)\rangle$ for the Pauli matrices $\hat{\sigma}_{x, y, z}$~\cite{SM}.
Therefore, we can calculate the Berry curvature $\mathcal{B}(k,t)=\left(\partial_k \tilde{\sigma}(k,t) \times \partial_\tau \tilde{\sigma}(k,t)\right) \cdot \tilde{\sigma}(k,t) / 2$ and the associated position shift in a modulation period
\begin{equation}
y(k,T_m)=-\frac{1}{d}\int_0^{T_m} \mathcal{B}(k,t) dt.
\end{equation}
When the system evolves along its instantaneous eigenstate, the average position shift
\begin{equation}
\bar{y}(T_m)=\frac{d}{2\pi}\int_0^{\frac{2\pi}{d}} y(k,T_m) dk,
\end{equation}
just gives the Chern number, that is, $\bar{y}(T_m)=C$.

\begin{figure}[htp]
\center
\includegraphics[width=\columnwidth]{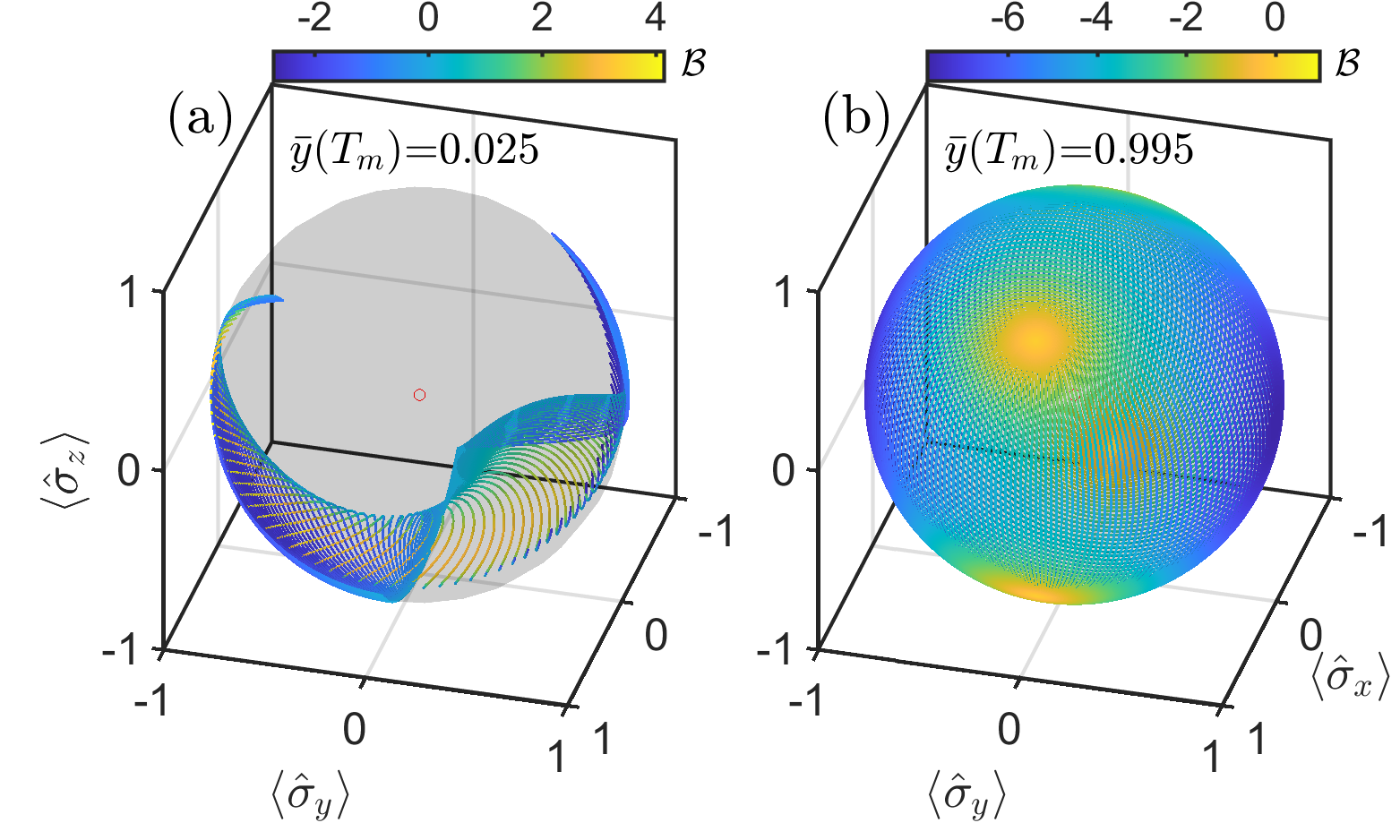}
\caption{The dynamical trajectories of $\left\langle\hat{\sigma}_{x, y, z}\right\rangle$ for two schemes: (a) Thouless pumping (TP) and (b) STA Thouless pumping (STATP).
The color denotes $\mathcal{B}=\left(\partial_k \tilde{\sigma} \times \partial_\tau \tilde{\sigma}\right) \cdot \tilde{\sigma} / 2$ calculated by viewing -$\left\langle\hat{\sigma}_{x, y, z}\right\rangle$ as an effective vector $\tilde{\sigma}$.
The average position shift $\bar{y}\left(T_m\right)$ as a two-dimensional integral in the $(k, t)$ parameter space are added into (a) and (b) with $\bar{y}\left(T_m\right)=0.025$ and $\bar{y}\left(T_m\right)=0.995$, respectively.
The parameters are chosen as $J=-1$, $\delta_{0}=0.8$, $\Delta_{0}=2$, $\omega=10$ and $\phi_{0}=0$.
The total evolution time is $T_{\mathrm{tot}}=T_m$.}
\label{fig:blochsphere}
\end{figure}

{\em Fast Thouless pumping under counterdiabatic control}. We compare two schemes to emphasize the fast Thouless pumping: (I) Thouless pumping for $\hat{H}_0(k, t)$ and (II) SAT Thouless pumping for $\hat{H}(k, t)$, respectively.
The parameters are chosen as $J=-1$, $\delta_{0}=0.8$, $\Delta_{0}=2$, $\phi_{0}=0$, and a common strong modulation frequency $\omega=10$.
Fig.~\ref{fig:blochsphere} shows the trajectories $\tilde{\sigma}(t)$ of the two schemes in the Bloch sphere expanded by $\left\langle\hat{\sigma}_{x, y, z}\right\rangle$, colored by $\mathcal{B}=\left(\partial_k \tilde{\sigma} \times \partial_\tau \tilde{\sigma}\right) \cdot \tilde{\sigma} / 2$.
The initial state involves all momemtum states for Thouless pumping and STA Thouless pumping schemes.
The trajectories of Thouless pumping only inhomogenously cover fraction of the Bloch sphere.
This is because the adiabatic condition fails in Thouless pumping.
When the driven frequency $\omega$ is comparable to or larger than the energy gap between two bands in Thouless pumping, the inevitable non-adiabatic transitions break quantization of band topology.
However, the trajectories of STA Thouless pumping can uniformly sample the Bloch sphere and rotate around the gap-closing point $(h^x_0=h^y_0=h^z_0=0)$ marked by red circles.
The controlled Hamiltonian significantly suppresses the transition between the instantaneous eigenstates of $\hat{H}_0(k, t)$, making the trajectories adiabatically follow the ones of instantaneous eigenstates, $\tilde{\sigma}(t)\approx \tilde{h}_0(t)$.
To characterize the degree of deviation from adiabaticity, we calculate the average position shift $\bar{y}\left(T_m\right)$ and position shift $y\left(k, T_m\right)$, and yield
$\bar{y}\left(T_m\right)=0.025$ for the breakdown of Thouless pumping and
$\bar{y}\left(T_m\right)=0.995 \approx C$ for STA Thouless pumping.
Attributed to the controlled Hamiltonian, the average position shift and the position shift in $T_m$ agree well with the Chern number in the STA Thouless pumping.

\begin{figure}[htp]
\center
\includegraphics[width=\columnwidth]{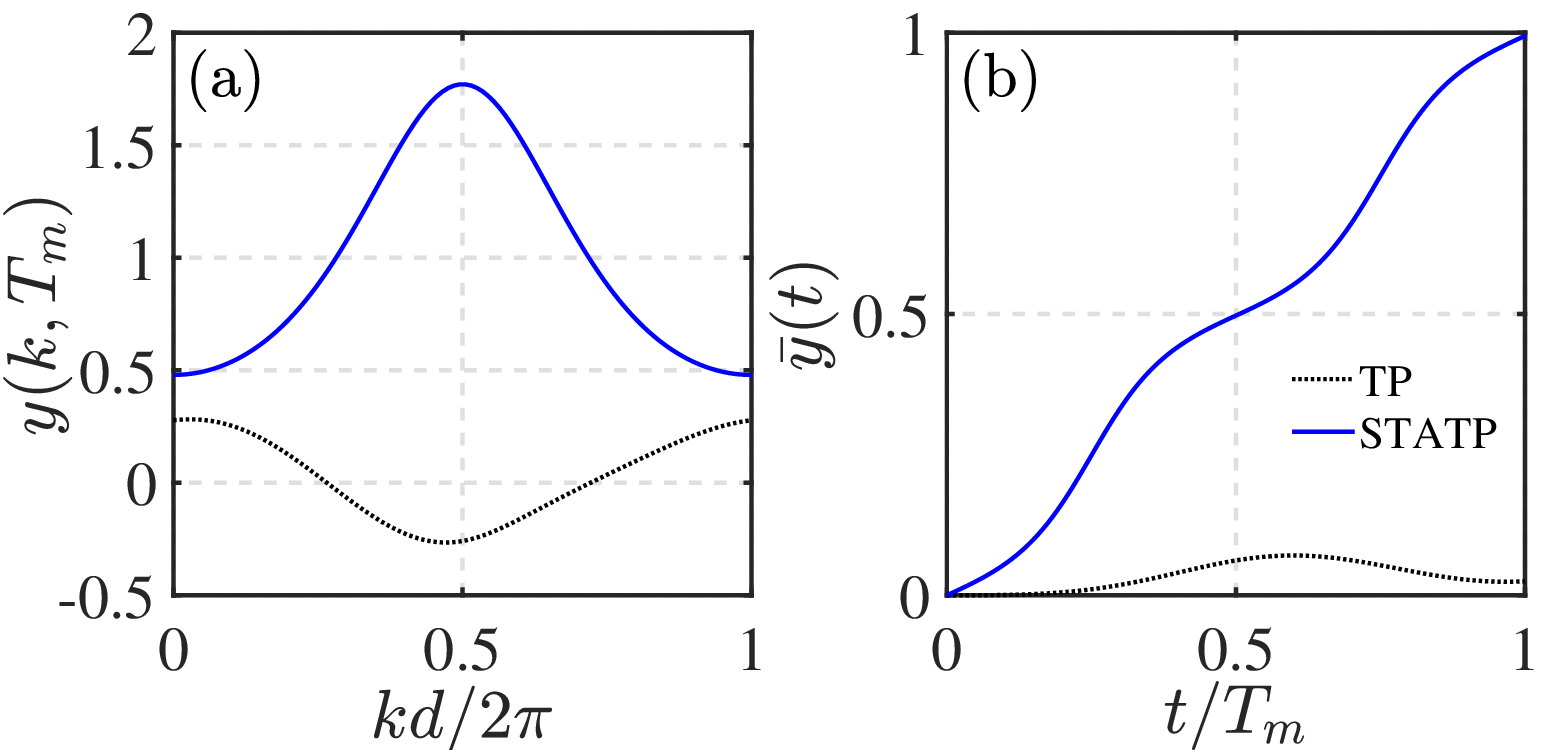}
\caption{(a) The position shift $y(k,T_m)$ over $T_m$ versus $k$ for two schemes: Thouless pumping (TP) with black dotted line and STA Thouless pumping (STATP) with blue solid line.
(b) The average position shift $\bar{y}(t)$ as a function of $t$ for Thouless pumping and STA Thouless pumping respectively.
The other parameters are chosen as the same as those in Fig.~\ref{fig:blochsphere}.}
\label{fig:cvsk}
\end{figure}

To further distinguish the two schemes, we calculate the position shift in $T_m$, and the (average) position shift as a function of time; see Figs.~\ref{fig:cvsk} (a) and (b), respectively.
The parameters are chosen the same as those in Fig.~\ref{fig:blochsphere}.
The black dotted and blue solid lines correspond to Thouless pumping and STA Thouless pumping, respectively.
For nonadiabatic breakdown, $y(k,T_m)$ oscillates around $0$ for Thouless pumping.
If one further averages $y(k,T_m)$ over each momentum, the average position shift $\bar{y}(t)$ does not approach and approaches to the Chern number in $T_m$, for Thouless pumping and STA Thouless pumping; see Fig.~\ref{fig:cvsk}(b).

\begin{figure}[htp]
\center
\includegraphics[width=\columnwidth]{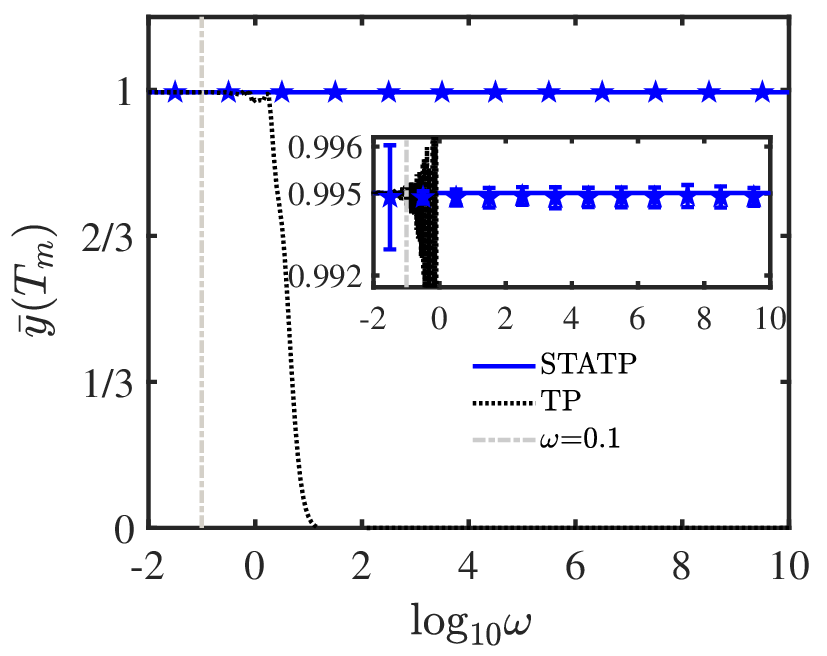}
\caption{The dependence of the average position shift $\bar{y}(T_m)$ [Thouless pumping (TP) and STA Thouless pumping (STATP)] on the modulation frequency $\omega$.
The gray dashed-dotted line is used to mark the adiabatic limit $\omega=10^{-1}$ where both the two schemes obtain quantized transports.
The error bars are captured by adding the noise of the coupling, indicating that our STATP with blue stars is robust against Hamiltonian fluctuations.
The other parameters are chosen as $J=-1$, $\delta_0=0.8$, $\Delta_0=2$ and $\phi_0=0$.
The total evolution time is $T_{\mathrm{tot}}=T_m$.}
\label{fig:cvsw}
\end{figure}

Whether the controlled Hamiltonian is applied or not, both schemes reach quantized transports
as long as $\hat{H}_0(k,t)$ is adiabatically driven with modulation frequency, that is, $\omega\leq 10^{-1}$ in Fig.~\ref{fig:cvsw}.
The controlled Hamiltonian contributes to fast quantized transport in a wide range of modulation frequency up to $\omega=10^{10}$ in our calculation results for STA Thouless pumping with a blue solid line in Fig.~\ref{fig:cvsw}.
Note that STA Thouless pumping can reach $10^{-11}$-fold increase of pumping efficiency compared to Thouless pumping.
To examine the robustness of the STA Thouless pumping, we consider the random Gaussian noise $\xi_t$ on the coupling term as $\hat{H}+\hat{H}_n$, where $\hat{H}_n$ is a non-diagonal matrix with $\hat{H}_n(1,2)=\hat{H}_n(2,1)=c \omega \xi_t$~\cite{SM}.
$c$ is the noise strength and $\left\langle\xi_t\right\rangle=0$ and $\left\langle\xi_t \xi_{t^{\prime}}\right\rangle=\delta\left(t-t^{\prime}\right)$.
For a fixed noise strength $c=0.5$, the small error bars in Fig.~\ref{fig:cvsw} show that the STA Thouless pumping with blue stars is robust against the moderate noise in the full range of modulation frequency.

{\em Fast Thouless pumping in position-space}. To further demonstrate the advantage of STA, we now pay attention to the two pumping schemes in the position space.
We obtain the real-space Hamiltonian $\hat{H}_R(t)$ by inverse Fourier transform of the bulk momentum-space Hamiltonian $\hat{H}(k, t)$.
The initial state is set as the Wannier state, which uniformly fills the lower band in Thouless pumping and STA Thouless pumping.
The evolved state $|\boldsymbol{\psi}(t)\rangle=\sum_{l} \psi_{l}(t)\left|l\right\rangle$ is determined by $|\boldsymbol{\psi}(t)\rangle=\mathcal{T} \exp \left\{-i \int_{0}^{t} \hat{H}_R(t)d t\right\}\left|\boldsymbol{\psi}\left(0\right)\right\rangle$
with the time-ordering operator $\mathcal{T}$.
We extract the mean displacement via
\begin{equation}
\Delta X(t)=X(t)-X(0),
\end{equation}
with  $X(t)=\sum_{l}l\left|\psi_{l}(t)\right|^{2}$,
and the change of wavepacket width
\begin{equation}
\Delta W(t)=W(t)-W(0)
\end{equation}
with wavepacket width described by $W(t)=\sqrt{\sum_{l}[l-X(t)]^2\left|\psi_{l}(t)\right|^2}$.
The change in wavepacket width is due to dynamical phase differences of different Bloch states~\cite{ke2016topological,PhysRevB.100.064302}.

\begin{figure}[!htp]
\center
\includegraphics[width=\columnwidth]{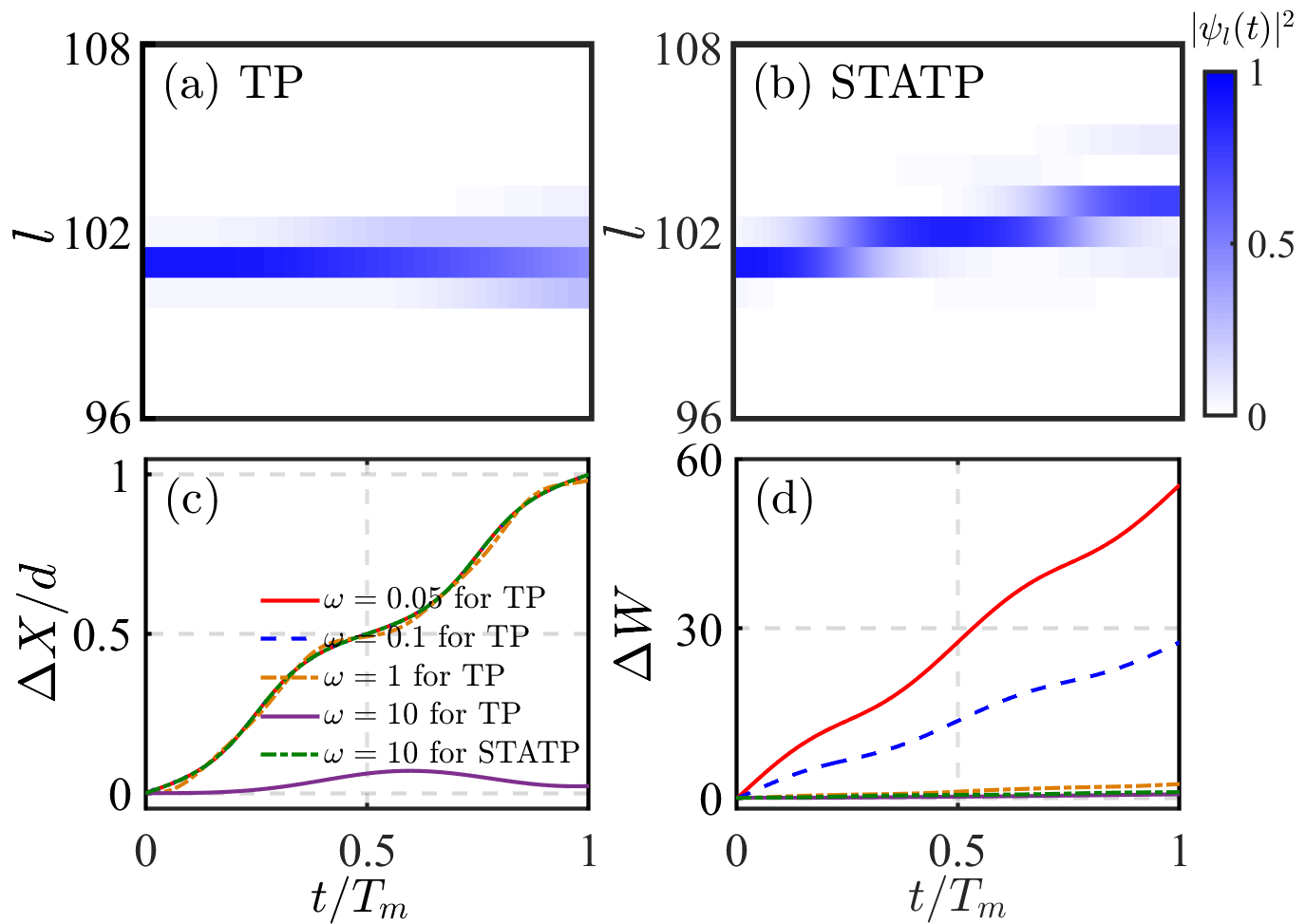}
\caption{The time evolution of position-space density distribution $|\psi_l(t)|^2$ for two schemes: (a) Thouless pumping (TP) and (b) STA Thouless pumping (STATP) with $\omega=10$.
(c) and (d) The center-of-mass displacement $\Delta X(t)/d$ and variation of wavepacket width $\Delta W(t)$ with different driven frequencies.
The other parameters are chosen as $J=-1$, $\delta_{0}=0.8$, $\Delta_{0}=2$ and $\phi_{0}=0$.}
\label{fig:realspace}
\end{figure}

On the one hand, STA Thouless pumping is successful and conventional Thouless pumping is broken down under a fixed high frequency of modulation.
Fig.~\ref{fig:realspace} shows the time evolution of density distribution $|\psi_l(t)|^2$, mean displacement and change of wavepacket width.
Fast modulation breaks the quantized transport of Thouless pumping due to the inevitable non-adiabatic effects; see Fig.~\ref{fig:realspace}(a) and the purple solid line in Fig.~\ref{fig:realspace}(c) with $\omega=10$.
However, even under fast modulation, the controlled Hamiltonian guarantees that the initial state evolves along the instantaneous eigenstates and is shifted by one unit cells in one cycles, which is determined by the Chern number; see Fig.~\ref{fig:realspace}(b) and green dashed-dotted line in Fig.~\ref{fig:realspace}(c) with $\omega=10$.
On the other hand, when both STA Thouless pumping and conventional Thouless pumping are successful, the STA Thouless pumping suffers less wavepacket diffusion than the conventional Thouless pumping.
Comparisons of displacement and change of width under different driven frequencies are presented in Figs.~\ref{fig:realspace}(c) and (d).
For Thouless pumping, as the driven frequency decreases, the displacement of the wavepacket is close to the quantized value but its width expands significantly.
The prominent wavepacket diffusion limits the practical applications of traditional Thouless pumping.
However, the faster modulation, the STA Thouless pumping suffers less wavepacket diffusion.
This is because less dynamical phase differences accumulate in a much shorter time for different Bloch states.
Even in the case of $\omega=10$, the change in wavepacket width is almost negligible; see the green dashed-dotted line in Fig.~\ref{fig:realspace}(d).
The technique of STA not only ensures quantized transport but suppresses the wavepacket dispersion in a shorter operator time.

{\em Conclusion and discussion}. In summary, we propose the use of Shortcuts to Adiabaticity (STA) to achieve Thouless pumping in both momentum and position spaces beyond the adiabatic regime.
We thoroughly investigate the robustness of our fast topological pumping when subjected to moderate Hamiltonian noise.
To validate the effectiveness of fast Thouless pumping through experimental means, it is essential to establish a well-defined experimental protocol~\cite{SM}.
In momentum space, we can implement our STA Thouless pumping Hamiltonian by connecting two levels of atoms through Raman transition~\cite{NC201671}.
The initial state can be prepared as a desired superposition of two hyperfine states by using external driving to induce coherent coupling.
The measurement of Pauli matrices at each time $t$ and momentum $k$ can be achieved by using cold atom techniques.
STA protocols allow for state manipulation within a significantly shorter timeframe compared to the decoherence time in cold atoms.

We believe that these results will motivate further studies on both sides of topological pumping and quantum control via STA.
First, the extension of STA topological pumping to higher-order~\cite{2022PhysRevB105195129,2022PhysRevLett128246602}, non-Abelian~\cite{2021PhysRevA103063518,2022PhysRevLett128244302,2022NP181080}, and fractional topological pumping scenarios~\cite{2023NP19420} adds an intriguing dimension to our exploration.
Because STA topological pumping is not limited by the energy gap, our protocol also provides a precise method for detecting topological phase transitions.
Second, as STA Thouless pumping is robust to random noise to some extent, we can also try to utilize a variety of topological phases to protect quantum control against random noise. Furthermore, since environment dissipation and incoherence are ubiquitous, it is worth exploring the role of topology in quantum control of open quantum systems.

\begin{acknowledgments}
We acknowledge useful discussions with Bo Lu and Chengyin Han.
This work is supported by the National Key Research and Development Program of China (Grant No. 2022YFA1404104), the National Natural Science Foundation of China (Grants No. 12025509 and 12275365), and the Natural Science Foundation of Guangdong Province (Grant No. 2023A1515012099).
\end{acknowledgments}

%


\onecolumngrid
\clearpage

\begin{center}
	\noindent\textbf{\large{Supplemental Material:}}
	\\\bigskip
	\noindent\textbf{\large{Shortcuts to adiabatic Thouless pumping}}
	\\\bigskip
	\onecolumngrid
	
	Wenjie Liu$^{1,2,3}$, Yongguan Ke$^{2,3,1,4,*}$, Chaohong Lee$^{2,3,1,\dag}$
	
	\small{$^1$ \emph{Institute of Quantum Precision Measurement, State Key Laboratory of Radio Frequency Heterogeneous Integration, Shenzhen University, Shenzhen 518060, China}}\\
	\small{$^2$ \emph{College of Physics and Optoelectronic Engineering, Shenzhen University, Shenzhen 518060, China}}\\
    \small{$^3$ \emph{Quantum Science Center of Guangdong-Hongkong-Macao Greater Bay Area (Guangdong), Shenzhen 518045, China}}\\
    \small{$^4$ \emph{Laboratory of Quantum Engineering and Quantum Metrology, School of Physics and Astronomy, Sun Yat-Sen University (Zhuhai Campus), Zhuhai 519082, China}}\\
\end{center}


\newcommand{\beginsupplement}{
    \setcounter{section}{0}
    \renewcommand{\thesection}{S\arabic{section}}
    \setcounter{equation}{0}
    \renewcommand{\theequation}{S\arabic{equation}}
    \setcounter{table}{0}
    \renewcommand{\thetable}{S\arabic{table}}
    \setcounter{figure}{0}
    \renewcommand{\thefigure}{S\arabic{figure}}
}

\beginsupplement

\section{Corresponding lattice Hamiltonian in position space} \label{realspacemodel}

Next we elaborate how to obtain the momentum-space Hamiltonian $\hat{H}_0(k, t)$ from a position-space bipartite lattice $\hat{H}_R(t)$.
Having been widely investigated, the position-space Rice-Mele model~\cite{1982PhysRevLett491455} is written as
\begin{equation}
\begin{aligned}
\hat{H}_R(t) & =\sum_{j=1}^{2 L}\left\{\left\{J+\delta_0 \sin [\pi j+\phi(t)]\right\} \hat{a}_j^{\dagger} \hat{a}_{j+1}+\text { H.c. }\right\} \\
& +\sum_{j=1}^{2 L}\Delta_0 \cos [\pi j+\phi(t)] \hat{n}_j
\end{aligned}
\end{equation}
where $\delta_0$ and $\Delta_0$ denote the modulation amplitudes of tunneling term and on-site potential, respectively.
The total system size is $2L$ with $L$ unit cells, and the length of each unit cell is $d=2a$.
For convenience, one can set $a=1$ and $\hbar=1$.
After employing the transformation relation
\begin{equation}\label{FFT}
\left\{\begin{array}{l}
\hat{a}_{2 j}^{\dagger}=\frac{1}{\sqrt{L}} \sum_k e^{i k 2 j} \hat{a}_{k, e}^{\dagger} \\
\hat{a}_{2 j-1}^{\dagger}=\frac{1}{\sqrt{L}} \sum_k e^{i k(2 j-1)} \hat{a}_{k, o}^{\dagger}
\end{array}\right.
\end{equation}
where $e$ ($o$) denotes the degree of freedom in even (odd) sublattice, the Hamiltonian in the quasimomentum space yields
\begin{equation}
\begin{aligned}
\hat{H}_0(k, t) & =\left[2 J \cos (k)-2 \delta_0 i \sin(k) \sin [\phi(t)]\right] \hat{a}_{k, e}^{\dagger} \hat{a}_{k, o}+\left[2 J \cos (k)+2 \delta_0 i \sin(k) \sin [\phi(t)]\right] \hat{a}_{k, o}^{\dagger} \hat{a}_{k, e} \\
& +\Delta_0 \cos [\phi(t)](\hat{a}_{k, e}^{\dagger} \hat{a}_{k, e}-\hat{a}_{k, o}^{\dagger} \hat{a}_{k, o}).
\end{aligned}
\end{equation}
When the $\hat{a}_{k, e}^{\dagger}$ ($\hat{a}_{k, o}^{\dagger}$) is regarded as spin up (down), the form of Hamiltonian in the quasimomentum space turns to be $\hat{H}_0(k, t)=\vec{h}_0 \cdot \vec{\sigma}$ with $\vec{h}_0$=($h^x_0$, $h^y_0$, $h^z_0$)=($2 J \cos (k) $, $2 \delta_0 \sin (k) \sin (\phi)$, $\Delta_0 \cos (\phi)$) and the Pauli matrices $\vec{\sigma}=(\hat{\sigma}_{x},\hat{\sigma}_{y},\hat{\sigma}_{z}$).
Therefore, the Hamiltonian (1) in quasimomentum space in the main text can be obtained in terms of the transformation relation~\eqref{FFT}.
There exists $\hat{H}_R(t)=\sum_k \hat{H}_0(k, t)$ with $k=\frac{\pi}{L} n$ ($n=0,1,2,...,L-1$).

\section{Derivation of the controlled Hamiltonian as well as its gauge invariance} \label{gaugeinvariance}

Here we clarify the derivation of Hamiltonian (4) in the main text and demonstrate its gauge invariance.
We consider a non-degenerate Hamiltonian~$\hat{H}_0(k, t)$~whose instantaneous eigenstates~$|n(t)\rangle$~and eigenvalues~$\varepsilon_n(t)$~satisfy~$\hat{H}_0(t)|n(t)\rangle=\varepsilon_n(t)|n(t)\rangle$.
Starting from an eigenstate $|n(0)\rangle$, the adiabatic theorem ensures the system remains in the $n$-th instantaneous eigenstate with $|\psi(t)\rangle=e^{i \beta(t)}|n(t)\rangle$.
$\beta(t)=\alpha(t)+\gamma(t)$ is the accumulated phase, including dynamical phase $\alpha(t)=-\frac{1}{\hbar} \int_0^t \varepsilon_n(\tau) d \tau$ and geometrical phase $\gamma(t)=i \int_0^t\left\langle n(\tau) \mid \partial_\tau n(\tau)\right\rangle d \tau$.
In the adiabatic approximation, the time-evolution operator can be described by
\begin{equation}\label{EvoultionOperator}
\hat{U}=\sum_n \exp \left[-\frac{i}{\hbar} \int_0^t \varepsilon_n(\tau) d \tau-\int_0^t\left\langle n(\tau) \mid \partial_\tau n(\tau)\right\rangle d \tau\right]|n(t)\rangle\langle n(0)|
\end{equation}
which obeys the Schr\"{o}dinger equation $i \hbar \frac{\partial}{\partial t} \hat{U}(t)=\hat{H}_0(t) \hat{U}(t)$.
Due to the fact that the controlled Hamiltonian $\hat{H}_C(k, t)$ aims at eliminating the transitions between instantaneous eigenstates
of $\hat{H}_0(k, t)$, it naturally has $i \hbar \frac{\partial}{\partial t} \hat{U}(t)=\left(\hat{H}_0(t)+\hat{H}_C(t)\right) \hat{U}(t)$.
Further, one can find
\begin{equation}\label{GeneralFormula}
\hat{H}_C(t)=i \hbar \frac{\partial U(t)}{\partial t} \hat{U}^{\dagger}(t)-\hat{H}_0(t).
\end{equation}
Substituting Eq.~\eqref{EvoultionOperator} into Eq.~\eqref{GeneralFormula}, an explicit form of the controlled Hamiltonian according to the counterdiabatic approach~\cite{JPhysChemB2005,Berry2009} becomes
\begin{equation}\label{GeneralFormula2}
\hat{H}_C(t)=i \hbar \sum_n\left|\partial_t n(t)\right\rangle\langle n(t)|-i \hbar \sum_n\left\langle n(t) \mid \partial_t n(t)\right\rangle| n(t)\rangle\langle n(t)|.
\end{equation}
Combing with the exactly solved eigenstates (2) of Hamiltonian (1) in the main text, its controlled Hamiltonian (4) in the main text is obtained according to Eq.~\eqref{GeneralFormula2}.

Once the gauge freedom is introduced, in general Eq.(2) in the main text is written as
\begin{equation}
\left\{\begin{array}{l}
\left|\lambda_{+}(t)\right\rangle=e^{i \delta_1(t)}\left(\begin{array}{c}
\cos \frac{\theta}{2} \\
\sin \frac{\theta}{2} e^{i \varphi}
\end{array}\right) \\
\left|\lambda_{-}(t)\right\rangle=e^{i \delta_2(t)}\left(\begin{array}{c}
-\sin \frac{\theta}{2} e^{-i \varphi} \\
\cos \frac{\theta}{2}
\end{array}\right)
\end{array}\right.
\end{equation}
as well as its derivation with respect to time $t$
\begin{equation}
\left\{\begin{aligned}
& \left|\partial_t \lambda_{+}\right\rangle=\mathrm{e}^{i \delta_1(t)} i \dot{\delta}_1(t)\left(\begin{array}{c}
\cos \frac{\theta}{2} \\
\sin \frac{\theta}{2} e^{i \varphi}
\end{array}\right)+\mathrm{e}^{i \delta_1(t)}\left(\begin{array}{l}
-\sin \frac{\theta}{2} \frac{\dot{\theta}}{2} \\
\cos \frac{\theta}{2} \frac{\dot{\theta}}{2} e^{i \varphi}+\sin \frac{\theta}{2} e^{i \varphi} i \dot{\varphi}
\end{array}\right) \\
& \left|\partial_t \lambda_{-}\right\rangle=\mathrm{e}^{i \delta_2(t)} i \dot{\delta}_2(t)\left(\begin{array}{l}
-\sin \frac{\theta}{2} e^{-i \varphi} \\
\cos \frac{\theta}{2}
\end{array}\right)+\mathrm{e}^{i \delta_2(t)}\left(\begin{array}{l}
-\cos \frac{\theta}{2} \frac{\dot{\theta}}{2} e^{-i \varphi}+\sin \frac{\theta}{2} e^{-i \varphi} i \dot{\varphi} \\
-\sin \frac{\theta}{2} \frac{\dot{\theta}}{2}
\end{array}\right)
\end{aligned}\right..
\end{equation}
After a detailed calculation, the controlled Hamiltonian keeps the same as the Hamiltonian (4) in the main text regardless of $\delta_1$ and $\delta_2$, which means the invariance for different choices of gauge.
The gauge invariance allows one to set $\delta_1=\delta_2=0$.

\section{Dynamical Berry curvature} \label{AdiabaticEvolution}

For an arbitrary state $|\psi(t)\rangle$, its density matrix reads
\begin{equation}\label{DensityMatrix}
\rho(t)=|\psi(t)\rangle\langle\psi(t)|=a_x \hat{\sigma}_x+a_y \hat{\sigma}_y+a_z \hat{\sigma}_z+\hat{a}_0 \hat{\sigma}_0.
\end{equation}
According to the orthogonality of Pauli matrices $\operatorname{Tr}\left(\hat{\sigma}_i \hat{\sigma}_j\right)=2 \delta_{i j}$, their expected values satisfy
\begin{equation}\label{AveragePauli}
\left\langle\hat{\sigma}_{x, y, z}\right\rangle=\operatorname{Tr}\left(\rho(t) \hat{\sigma}_{x, y, z}\right)=2 a_{x, y, z}.
\end{equation}
One can describe $\hat{H}_0(k, t)$ in terms of its instantaneous eigenbasis by
\begin{equation}
\hat{H}_0(k,t)=\varepsilon_{+}(t)\left|\lambda_{+}(t)\right\rangle\left\langle\lambda_{+}(t)\left|-\varepsilon_{+}(t)\right| \lambda_{-}(t)\right\rangle\left\langle\lambda_{-}(t)\right|
\end{equation}
with $\varepsilon_{+}=\sqrt{(h^x_0)^2+(h^y_0)^2+(h^z_0)^2}$.
Using identity $\sum_{x=\pm}\left|\lambda_x(t)\right\rangle\left\langle\lambda_x(t)\right|=I$, it is easy to find \begin{equation}
\left|\lambda_{-}(t)\right\rangle\left\langle\lambda_{-}(t)\right|=\frac{\varepsilon_{+}(t) I-\hat{H}_0(k,t)}{2 \varepsilon_{+}(t)}.
\end{equation}
When the controlled Hamiltonian is applied, the system stays at the instantaneous eigenstate $\left|\lambda_{-}(t)\right\rangle$ from the chosen initial state $\left|\lambda_{-}(0)\right\rangle$, and it subsequently yields
\begin{equation}\label{InstantaState}
\rho(t)=|\psi(t)\rangle\left\langle\psi(t)|=| \lambda_{-}(t)\right\rangle\left\langle\lambda_{-}(t)\right|=\frac{\varepsilon_{+}(t) I-\hat{H}_0(k,t)}{2 \varepsilon_{+}(t)}.
\end{equation}
Then it has $h^{x, y, z}_0=-2\varepsilon_{+}a_{x,y,z}$.
In terms of Eq.~\eqref{DensityMatrix}, Eq.~\eqref{AveragePauli} and Eq.~\eqref{InstantaState}, we obtain $h^{x, y, z}_0=-\varepsilon_{+}\left\langle\hat{\sigma}_{x, y, z}\right\rangle$.
Until to now, we has demonstrated
\begin{equation}\label{DynamicalBerry}
\tilde{h}_0=\vec{h}_0 /|\vec{h}_0|=\left(-\left\langle\hat{\sigma}_x\right\rangle,-\left\langle\hat{\sigma}_y\right\rangle,-\left\langle\hat{\sigma}_z\right\rangle\right).
\end{equation}
Therefore, the Berry curvature can be dynamically acquired from Eq.~\eqref{DynamicalBerry}.

\section{Experimental feasibility} \label{Experiment}

\begin{figure}[htp]
\center
\includegraphics[width=0.5\columnwidth]{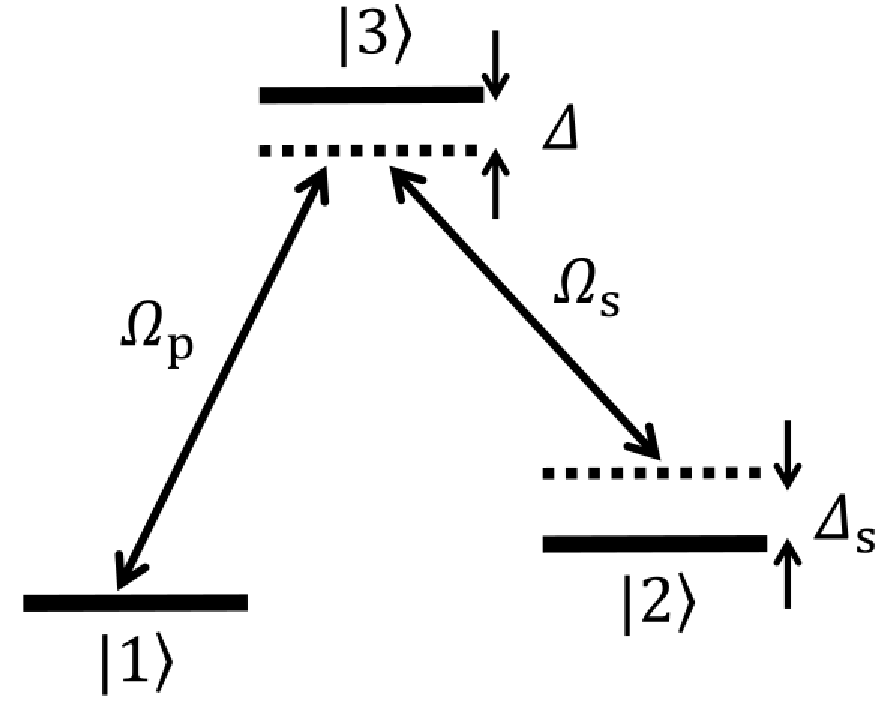}
\caption{Two ground states $|2\rangle$, $|1\rangle$ and one excited state $|3\rangle$ consist of a typical three-level $\Lambda$ structure.
The pumping laser $\Omega_p(t)$ (Stokes laser $\Omega_s(t)$) is used to couple the ground state $|1\rangle$ ($|2\rangle$ ) with the excited state $|3\rangle$.
$\Delta$ characterizes the single-photon detuning which approaches to 2$\pi$$\times$2.5GHz for ${ }^{87} \mathrm{Rb}$ atoms.
The two-photon detuning is selected as $\Delta_s=0$.
}
\label{fig:threelevel}
\end{figure}

The technique of shortcuts to adiabaticity is proposed to speed up the typical two- and three-level adiabatic passage~\cite{2010PhysRevLett105123003}.
The stimulated Raman passage has been experimentally demonstrated in ultracold atoms~\cite{NC201671} whose original
system is given by
\begin{equation}\label{originalthreelevel}
\begin{aligned}
H(t) & =\omega_3|3\rangle\left\langle 3\left|+\omega_2\right| 2\right\rangle\left\langle 2\left|+ \omega_1\right| 1\right\rangle\langle 1| \\
& +\Omega_p(t) \cos \left(\omega_p t+\phi_p\right)|3\rangle\left\langle 1\left|+\Omega_s(t) \cos \left(\omega_s t+\phi_s\right)\right| 3\right\rangle\langle 2|+\text { H.c. }
\end{aligned}.
\end{equation}
The corresponding three-level $\Lambda$ structure of ${ }^{87} \mathrm{Rb}$ atoms is presented in Fig.~\ref{fig:threelevel} with the Rabi frequencies for the Stokes laser $\Omega_s(t)$ and pumping laser $\Omega_p(t)$.
Implementing the rotating-wave approximation, Eq.~\eqref{originalthreelevel} turns to be
\begin{equation}
\begin{aligned}
H(t) & =\omega_3|3\rangle\left\langle 3\left|+\omega_2\right| 2\right\rangle\left\langle 2\left|+ \omega_1\right| 1\right\rangle\langle 1| \\
& +\frac{\Omega_p(t)}{2} e^{-i \left(\omega_p t+\phi_p\right)}|3\rangle \langle 1|+\frac{\Omega_s(t)}{2} e^{-i \left(\omega_s t+\phi_s\right)}
| 3 \rangle\langle 2|+\text { H.c. }
\end{aligned}.
\end{equation}
For the two-photon resonance case ($\Delta_s=0$), the system is transformed in the rotating frame as
\begin{equation}\label{threelevelstru}
H^I(t)=\frac{1}{2}\left(\begin{array}{ccc}
2 \Delta & \Omega_s(t) & \Omega_p(t) e^{-i \phi_L} \\
\Omega_s(t) & 0 & 0 \\
\Omega_p(t) e^{i \phi_L} & 0 & 0
\end{array}\right)
\end{equation}
in the basis ($|3\rangle,|2\rangle,|1\rangle$) and the time-dependent phase is removed.
Here $\Delta=\omega_3-\omega_1-\omega_p=\omega_3-\omega_2-\omega_s$ is single-photon detuning and $\phi_L=\phi_p-\phi_s$ denotes phase difference between two coupling lasers.
Under the condition of large detuning $\Delta \gg \Omega_p, \Omega_s$, the population on the level $|3\rangle$ remains unchanged with $\dot{c}_3(t) \simeq 0$.
$H^I(t)$ can be divided into the dominant part
\begin{equation}\label{threelevelstru}
H_0^I(t)=\frac{1}{2}\left(\begin{array}{ccc}
2 \Delta & 0 & 0 \\
0 & 0 & 0 \\
0 & 0 & 0
\end{array}\right)
\end{equation}
and the perturbation part
\begin{equation}\label{threelevelstru}
H_1^I(t)=\frac{1}{2}\left(\begin{array}{ccc}
0 & \Omega_s(t) & \Omega_p(t) e^{-i \phi_L} \\
\Omega_s(t) & 0 & 0 \\
\Omega_p(t) e^{i \phi_L} & 0 & 0
\end{array}\right).
\end{equation}
$H^I(t)$ becomes $e^{\hat{S}} H^I(t) e^{-\hat{S}}$ via an anti-Hermitian matrix $\hat{S}$.
Subsequently, we have
\begin{equation}
\begin{aligned}
& e^{\hat{S}} H_0^I(t) e^{-\hat{S}}=H_0^I(t)+\frac{1}{1 !}\left[\hat{S}, H_0^I(t)\right]+\frac{1}{2 !}\left[\hat{S},\left[\hat{S}, H_0^I(t)\right]\right]+\ldots \\
& e^{\hat{S}} H_1^I(t) e^{-\hat{S}}=H_1^I(t)+\frac{1}{1 !}\left[\hat{S}, H_1^I(t)\right]+\ldots
\end{aligned}.
\end{equation}
Assuming that $H_1^I(t)+\left[\hat{S}, H_0^I(t)\right]=0$, the perturbation expansion up to the second order reads
\begin{equation}\label{secondorder}
H^I(t) \approx H_0^I(t)+\frac{1}{2 !}\left[\hat{S}, H_1^I(t)\right].
\end{equation}
In order to satisfy the condition $H_1^I(t)+\left[\hat{S}, H_0^I(t)\right]=0$, the anti-Hermitian matrix is determined as
\begin{equation}\label{antiHemer}
\hat{S}=\frac{\Omega_s}{2 \Delta}|3\rangle\langle 2|+\frac{\Omega_p e^{-i \phi_L}}{2 \Delta}| 3\rangle\langle 1|-\frac{\Omega_s}{2 \Delta}| 2\rangle\langle 3|-\frac{\Omega_p e^{i \phi_L}}{2 \Delta}| 1\rangle\langle 3|.
\end{equation}
When substituting Eq.~\eqref{antiHemer} into Eq.~\eqref{secondorder}, after adiabatically  eliminated the three-level system~\eqref{threelevelstru} can be reduced to an effective two-level model in the basis ($|2\rangle,|1\rangle$)
\begin{equation}\label{efftwolevel}
H_{\text {eff }}=\frac{1}{2}\left(\begin{array}{cc}
\Delta_{\text {eff }}(t) & \Omega_{\text {eff }}(t) e^{-i \phi_L} \\
\Omega_{\text {eff }}(t) e^{i \phi_L} & -\Delta_{\text {eff }}(t)
\end{array}\right)
\end{equation}
with $\Delta_{\text {eff }}(t)=\frac{\Omega_p^2(t)-\Omega_s^2(t)}{4 \Delta}$ and $\Omega_{\text {eff }}(t)=-\frac{\Omega_p(t) \Omega_s(t)}{2 \Delta}$.

\begin{figure}[htp]
\center
\includegraphics[width=0.9\columnwidth]{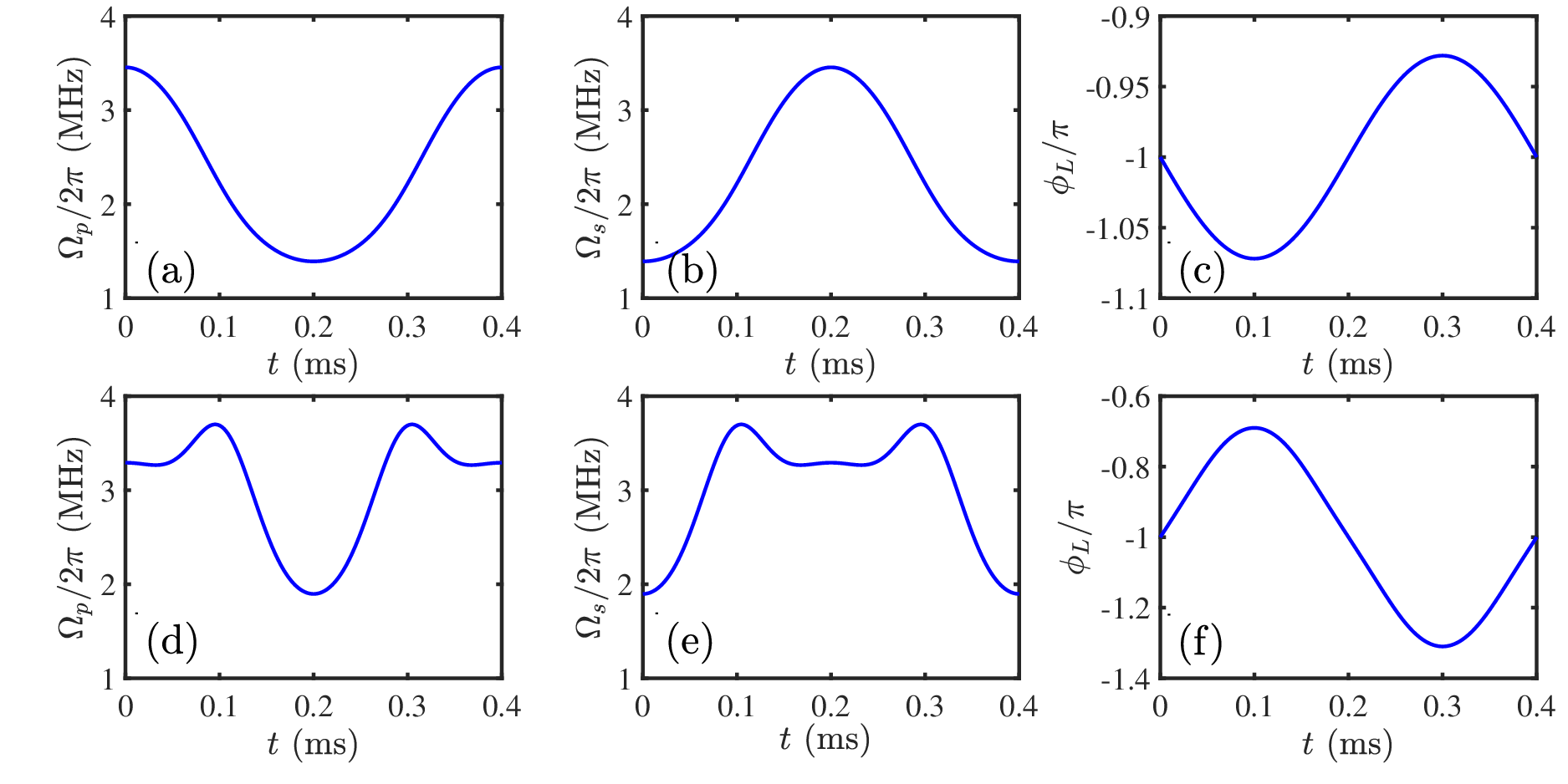}
\caption{(a)-(c) The time sequence of the original coupling lasers obtained from the original Rice-Mele model.
(d)-(f) The modified coupling lasers are equivalent to add the controlled Hamiltonian into the original Rice-Mele model.
The single-photon detuning is set as $\Delta$=2$\pi$$\times$2.5GHz for ${ }^{87} \mathrm{Rb}$ atoms.
The other parameters are chosen as $J/\Delta=-1\times10^{-7}$, $\delta_{0}/\Delta=0.8\times10^{-7}$, $\Delta_{0}/\Delta=2\times10^{-7}$, $\omega/\Delta=10\times10^{-7}$ and $\phi_{0}=0$.
The total evolution time is $T_m=0.4$ms.}
\label{fig:couplinglaser10}
\end{figure}

\begin{figure}[htp]
\center
\includegraphics[width=0.9\columnwidth]{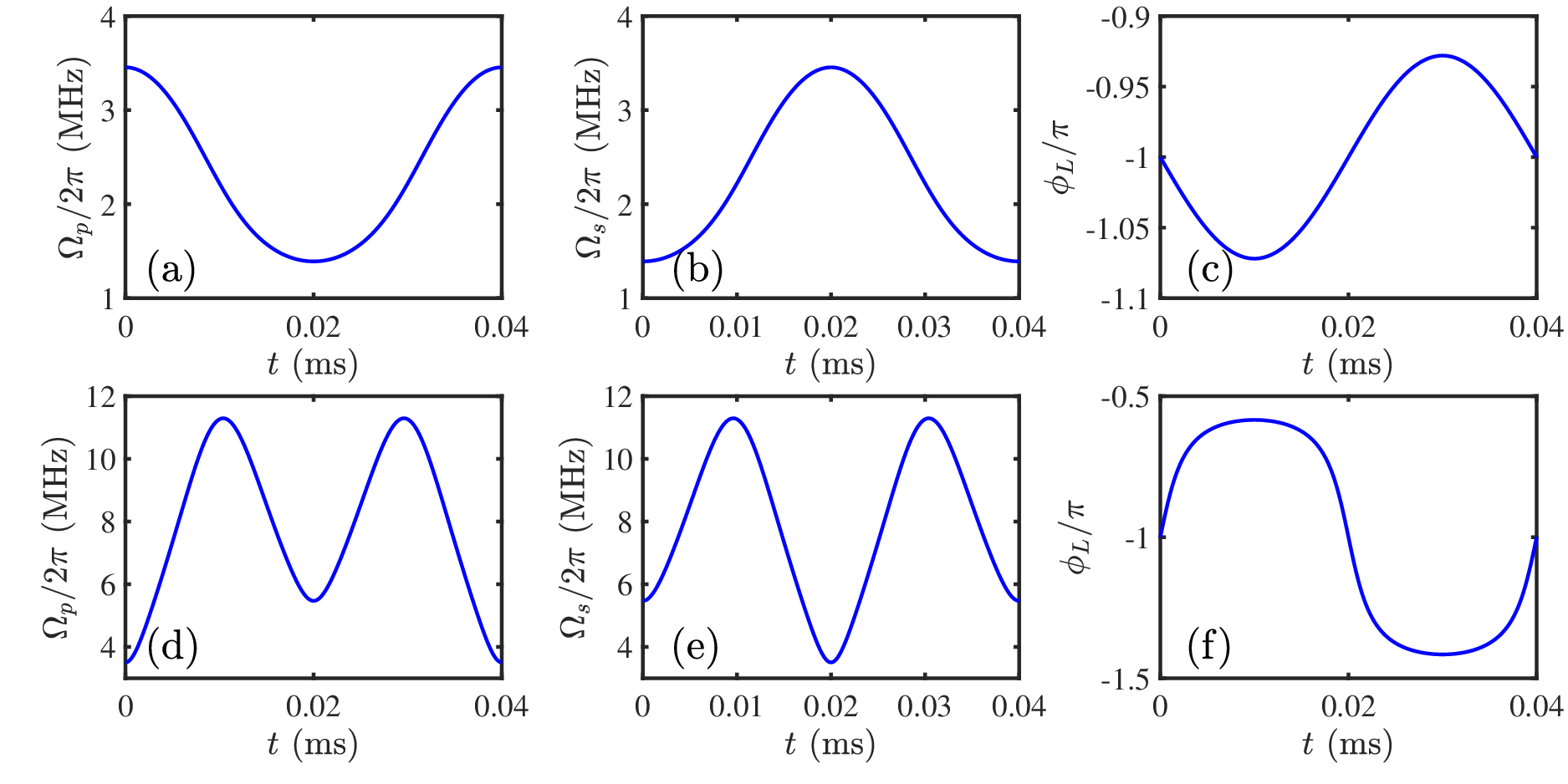}
\caption{(a)-(c) The time sequence of the original coupling lasers obtained from the original Rice-Mele model.
(d)-(f) The modified coupling lasers are equivalent to add the controlled Hamiltonian into the original Rice-Mele model.
The parameters are the same with Fig.~\ref{fig:couplinglaser10} except for $\omega/\Delta=100\times10^{-7}$.
The total evolution time is $T_m=0.04$ms.}
\label{fig:couplinglaser100}
\end{figure}

Equivalently, our two-level Rice-Mele model can be written as
\begin{equation}\label{twolevelRM}
\hat{H}_0(k, t)=\left(\begin{array}{cc}
h_0^z & h_0^x-i h_0^y \\
h_0^x+i h_0^y & -h_0^z
\end{array}\right)=\left(\begin{array}{cc}
h_0^z & \sqrt{\left(h_0^x\right)^2+\left(h_0^y\right)^2} e^{i \phi_h} \\
\sqrt{\left(h_0^x\right)^2+\left(h_0^y\right)^2} e^{-i \phi_h} & -h_0^z
\end{array}\right)
\end{equation}
with $\phi_h$=arctan$[-h_0^y/h_0^x]$.
Combining Eq.~\eqref{efftwolevel} and Eq.~\eqref{twolevelRM}, there exist the following relations
\begin{equation}
\left\{\begin{array}{l}
h_0^z=\frac{\Delta_{\text {eff }}}{2}=\frac{\Omega_p^2-\Omega_s^2}{8 \Delta} \\
\sqrt{\left(h_0^x\right)^2+\left(h_0^y\right)^2}=\frac{\Omega_{\text {eff }}}{2}=-\frac{\Omega_p \Omega_s}{4 \Delta} \\
\phi_h=-\phi_L
\end{array}\right..
\end{equation}
If the single-photon detuning is a fixed value ($\Delta$=2$\pi$$\times$2.5GHz for ${ }^{87} \mathrm{Rb}$ atoms), it is easy to
find that our target Hamiltonian can be achieved by engineering the coupling lasers as:
\begin{equation}
\begin{aligned}
& \Omega_s^2=-4 \Delta h_0^z+4 \Delta \sqrt{\left(h_0^z\right)^2+\left(h_0^x\right)^2+\left(h_0^y\right)^2} \\
& \Omega_p^2=16 \Delta^2\left[\left(h_0^x\right)^2+\left(h_0^y\right)^2\right] /\left(-4 \Delta h_0^z+4 \Delta \sqrt{\left(h_0^z\right)^2+\left(h_0^x\right)^2+\left(h_0^y\right)^2}\right)
\end{aligned}.
\end{equation}
When $\omega/\Delta=10\times10^{-7}$, we fix a value of initial quasimomentum $k$,
and the coupling lasers for the original Rice-Mele model without the controlled Hamiltonian are exhibited in Figs.~\ref{fig:couplinglaser10}(a)-(c).
In order to absorb the controlled Hamiltonian into the
original Rice-Mele model, the coupling lasers are modified as shown in Figs.~\ref{fig:couplinglaser10}(d)-(f).
We note that even for a high-frequency driving the system evolves along its eigenstate by modifying the shapes of coupling lasers.
Keeping the other parameters unchanged, the shapes with $\omega/\Delta=100\times10^{-7}$ change more significantly than the case with $\omega/\Delta=10\times10^{-7}$ but for a shorter operation time, as shown in Figs.~\ref{fig:couplinglaser100}(d)-(f).
The coupling lasers satisfy the large detuning whether the controlled Hamiltonian is applied or not.
The operator time locates on the regime of ${ }^{87} \mathrm{Rb}$-atom experiments.
The abundant experimental technique ensures the accurate detection of Pauli matrices.
Then the initial quasimomentum samples the whole Brillouin zone, the expectation values of Pauli matrices at each moment and quasimomemtum contribute to the average position shift $\bar{y}\left(T_m\right)$ as a two-dimensional integral in the $(k, t)$ parameter space.

The experimental imperfection originates from the strength fluctuations of the coupling lasers and single-photon detuning,
and leads to two kinds of noises including the fluctuation of the non-diagonal term
\begin{equation}
\hat{H}_n=c \omega \xi_t\left(\begin{array}{ll}
0 & 1 \\
1 & 0
\end{array}\right)
\end{equation}
and
the fluctuation of diagonal term
\begin{equation}
\hat{H}_n=c \omega \xi_t\left(\begin{array}{ll}
1 & 0 \\
0 & -1
\end{array}\right).
\end{equation}
$c$ is the fluctuation strength and $\left\langle\xi_t\right\rangle=0$ and $\left\langle\xi_t \xi_{t^{\prime}}\right\rangle=\delta\left(t-t^{\prime}\right)$.
The robustness of STA Thouless pumping has been demonstrated against the experimental imperfections in Fig.3 in the main text.
The resulting robustness to diagonal term is similar with the non-diagonal one.
Therefore our STA Thouless pumping is feasible in cold-atom experiments.


\begin{thebibliography}{51}
\bibitem [{SM()}]{SM}%
  \BibitemOpen
  \bibfield  {title} {\enquote {\bibinfo {title} {See {S}upplemental {M}aterial
  for details of ({I}) {C}orresponding lattice {H}amiltonian in position space;
  ({II}) {D}erivation of the control {H}amiltonian as well as its gauge
  invariance; ({III}) {D}ynamical {B}erry curvature; ({IV}) {E}xperimental
  feasibility},}\ }\href@noop {} {\ }\BibitemShut {NoStop}%
\end{thebibliography}
\end{document}